\documentclass[11pt]{article}

\usepackage[margin=1in]{geometry}
\usepackage[T1]{fontenc}
\usepackage[utf8]{inputenc}
\usepackage{lmodern}
\usepackage{microtype}
\usepackage{amsmath,amssymb,amsfonts}
\usepackage{booktabs}
\usepackage{longtable}
\usepackage{graphicx}
\usepackage{xcolor}
\usepackage{hyperref}
\usepackage{caption}
\usepackage{subcaption}
\usepackage[numbers,sort&compress]{natbib}

\hypersetup{
  colorlinks=true,
  linkcolor=blue!50!black,
  citecolor=blue!50!black,
  urlcolor=blue!50!black
}

\newcommand{\method}{\textsc{Conservative FluxNet}}
\newcommand{\baseline}{\textsc{Direct StateNet}}
\newcommand{\manufig}[1]{figures/#1}

\title{Conservative Discrete Structure Stabilizes Autoregressive Rollouts in a 1D Drift Diffusion Poisson Benchmark}
\author{Yufeng Wang*\\
\small Stony Brook University
\and
Lu Wei*\\
\small Stony Brook University
\and
Haibin Ling$\dag$\\
\small Westlake University
}
\begin{document}
\maketitle

% Add footnote for coauthors
\begingroup
\renewcommand\thefootnote{}
\footnotetext{* These authors contributed equally to this work.}
\footnotetext{$\dag$ Corresponding author.}
\endgroup

\begin{abstract}
Learned plasma transport surrogates can match short horizon states while failing over long rollouts because charge accounting, density admissibility, and Poisson compatible field reconstruction are not enforced. We study this issue in a controlled nondimensional one dimensional drift diffusion Poisson benchmark with Dirichlet electrostatic potential boundaries and zero species wall fluxes. The benchmark is a conservation and rollout test, not a complete sheath wall model. We compare \method, a structure preserving flux correction model with a conservative finite volume update and positivity aware limiting, against direct next state regressors, direct variants with Poisson recomputation, charge projection, and rollout training, and a classical conservative core without learned correction. The central result is that the classical finite volume core alone achieves near roundoff rollout error, so the paper is primarily about conservative discrete structure rather than learned closure. On the headline experiment, the conservative model achieves rollout MSE $7.35\times 10^{-9}$ versus $4.23\times 10^{1}$ for the unconstrained baseline, $2.53\times 10^{1}$ with Poisson recomputation, $6.72\times 10^{1}$ with charge projection, and $2.71\times 10^{1}$ with four step rollout training. Across $64$ prespecified configurations, it wins rollout mean squared error in $60/64$ cases despite winning one step mean squared error in only $19/64$. These results show that, for this controlled benchmark and comparison class, local conservative finite volume structure is more important than one step neural regression accuracy for stable autoregressive rollout.
\end{abstract}

\section{Introduction}
Neural surrogates for time dependent partial differential equations (PDEs) are attractive because they can amortize repeated simulation costs and provide differentiable approximations to expensive solvers. This promise has driven rapid progress in physics informed neural networks (PINNs), neural operators, and hybrid differentiable solvers \citep{raissi2019pinn,li2021fno,kovachki2023neuraloperator,li2021pino}. However, many scientific simulations are not judged by one step interpolation error. They are judged by whether long rollouts preserve the physical budgets, signs, couplings, and boundary effects that determine whether a predicted trajectory remains meaningful.

Plasma transport is a sharp example of this difficulty. Drift Diffusion Poisson (DDP) fluid models are a standard reduced description for low temperature plasma transport: they have been used for decades in discharge modeling, plasma processing, and sheath analysis, with seminal treatments in Lieberman and Lichtenberg \citep{lieberman2005principles}, Hagelaar and Pitchford \citep{hagelaar2005boltzmann}, and conservative discretization theory tracing to Scharfetter and Gummel \citep{scharfetter1969large}. In DDP models, charged species are transported by diffusion and by electric fields that are themselves generated by the charge density. A small density error can change the electric field, the changed field can alter subsequent transport, and the resulting feedback can accumulate over many steps. Full sheath physics additionally involves wall collection, emission, sheath edge constraints, and kinetic effects \citep{lieberman2005principles,riemann1991sheath,hagelaar2005boltzmann}. The benchmark studied here does not attempt to model all of that physics. Instead, it isolates the numerical surrogate learning question that appears inside such models: can a learned update preserve charge accounting, density admissibility, and Poisson compatibility during long autoregressive rollout?

Existing physics informed approaches address this problem only partially. PINNs and related methods encode equations through residual penalties, but residual minimization does not by itself guarantee that a time stepping map will conserve a discrete invariant over long rollout \citep{raissi2019pinn,basir2023auglag,deryck2024preconditioning}. Neural operators learn maps between function spaces and have shown strong performance for parametric PDEs, but a generic operator architecture does not automatically inherit finite volume cancellation, positivity preservation, or Poisson compatible field reconstruction \citep{li2021fno,kovachki2023neuraloperator,li2021pino}. Recent conservative and structure preserving neural PDE methods move closer to the needed design principle by learning fluxes, corrections, limiters, or invariant preserving projections rather than replacing the full solver \citep{chen2024learning,lichtle2025nfv,huang2025tvd,cardoso2024exactly,liu2025correction,liu2025entropy,shaffer2026structure}.

This paper asks a specific question: for transient drift diffusion Poisson plasma transport, is long horizon physical fidelity improved by putting prediction inside a conservative update rather than asking a neural network to predict the next state directly? We study this question on a controlled one dimensional benchmark. The benchmark is intentionally simpler than a full plasma wall model: it uses Dirichlet electrostatic potential boundaries and zero species wall fluxes. This simplification isolates the numerical question of interest: whether a plasma surrogate can maintain charge accounting, density admissibility, and stable rollout when the governing transport structure is built into the update map.

We compare two designs. The first, \baseline, directly predicts the next state from the current state. The second, \method, keeps the update in conservative flux form, represents electrostatic potential on nodes, computes electric fields by discrete differentiation, and learns only correction terms on the transport path. Because global charge cancellation is algebraic in the conservative update, near roundoff charge error should be interpreted as an enforced structural property rather than as a discovered neural behavior. The empirical question is whether that structural property also improves state rollout accuracy and admissibility under the tested data regimes.

The contributions of this paper are modest but specific. First, we formulate a compatible drift diffusion Poisson rollout model with shared face conservative updates, Poisson compatible field reconstruction, and positivity aware flux limiting. Second, we introduce a benchmark protocol that evaluates one step accuracy together with rollout error, charge drift, and density admissibility across seed repeats, stress tests, generalization shifts, capacity changes, and loss weight studies. Third, we show empirically that physical fidelity metrics can contradict one step error rankings in this benchmark, and that these metrics are essential for interpreting learned plasma transport models.

\section{Related Work}
\subsection{Physics Informed and Operator Learning Surrogates}
PINNs introduced a flexible framework for solving forward and inverse PDE problems by penalizing governing equation residuals in the training objective \citep{raissi2019pinn}. Subsequent physics informed machine learning work broadened this idea to multi physics modeling, inverse problems, and scientific discovery. These methods are attractive because they can incorporate measurements, constraints, and differential equations in a single optimization problem. Their limitation for the present setting is that a small residual at sampled points is not the same as an exact discrete conservation law under repeated time stepping. Training pathologies and constraint balancing problems have motivated augmented Lagrangian and preconditioning approaches \citep{basir2023auglag,deryck2024preconditioning}, but these still act through optimization rather than through the algebra of the update.

Neural operators address a different bottleneck by learning solution operators rather than single instance solutions. Fourier neural operators and related architectures provide mesh flexible maps for parametric PDE families \citep{li2021fno,kovachki2023neuraloperator}, and physics informed neural operators incorporate residual information into operator learning \citep{li2021pino}. These methods can be highly effective for interpolation and parametric generalization, but generic operator layers do not automatically enforce species conservation, positivity, or electrostatic compatibility. For plasma transport, those properties are not optional diagnostics; they are part of what makes a rollout interpretable.

\subsection{Structure Preserving Neural PDE Solvers}
A growing line of work addresses the above limitation by embedding numerical structure directly into neural PDE models. Conservative form networks learn fluxes for conservation laws rather than unconstrained state maps \citep{chen2024learning}. Neural finite volume methods preserve the finite volume control volume viewpoint while learning closures or flux components \citep{lichtle2025nfv}. Learned limiter approaches keep the solver form fixed and learn the stabilization component \citep{huang2025tvd}, while entropy stable and correction layer methods add stronger invariant or stability control \citep{liu2025entropy,liu2025correction}. Exactly conservative PINN and DeepONet variants similarly emphasize that global invariants should be built into the architecture or corrected explicitly rather than left only to a penalty term \citep{cardoso2024exactly}.

Our method follows this structure preserving direction but specializes it to a Poisson coupled plasma transport problem. The key distinction is that conservation is not the only structural requirement. The model must also keep the potential field relation compatible, respect species density admissibility, and evaluate performance over long rollouts where electrostatic feedback can amplify small local errors.

\begin{table}[t]
  \centering
  \caption{Relationship between the present benchmark and common learned PDE method classes. The entries summarize the properties targeted by the method class itself, not all possible variants. ``Partial'' means the property can be encouraged or added through losses or projections, but is not generally enforced by the base architecture. DDP denotes drift diffusion Poisson.}
  \label{tab:method_class_comparison}
  \resizebox{\linewidth}{!}{
  \begin{tabular}{lccccc}
    \toprule
    Method class & Exact charge accounting & Positivity handling & Poisson compatibility & Learns closure & DDP plasma test \\
    \midrule
    Residual PINN & Partial & Partial & Partial & Yes & Possible \\
    Generic neural operator & No & No & No & Yes & Possible \\
    Conservation projected state model & Global only & Partial & Optional & Yes & Not tested here \\
    Neural finite volume method & Yes & Optional & Optional & Yes & Usually not DDP specific \\
    Classical finite volume solver & Yes & Scheme dependent & Yes & No & Yes \\
    \method & Yes & Flux limited & Yes & Bounded flux correction & Yes, controlled 1D benchmark \\
    \bottomrule
  \end{tabular}
  }
\end{table}

\subsection{Plasma Transport and Drift Diffusion Poisson Models}
Fluid plasma models based on continuity equations, Drift Diffusion fluxes, and Poisson coupling are standard reduced descriptions for low temperature plasma transport and discharge modeling \citep{lieberman2005principles,hagelaar2005boltzmann}. Sheath formation and sheath edge constraints are central to plasma wall interaction theory \citep{riemann1991sheath,baalrud2011bohm}. These models are less expensive than kinetic descriptions, but they are still numerically delicate because density, electric field, and boundary behavior are tightly coupled.

Classical Drift Diffusion discretizations have long used conservative finite volume or exponentially fitted ideas to control transport and preserve physically meaningful flux balances \citep{scharfetter1969large,bessemoulin2012finite}. This history motivates the design choice in the present paper: a learned model should not discard the conservative bookkeeping that makes the classical solver reliable. Instead, learning should enter through correction terms or closures while the update remains a conservative transport step.

\section{Preliminaries and Problem Setup}
\begin{table}[t]
  \centering
  \caption{Core notation used in the benchmark and evaluation.}
  \label{tab:nomenclature}
  \begin{tabular}{ll}
    \toprule
    Symbol & Meaning \\
    \midrule
    $n_e, n_i$ & electron and ion cell averaged densities \\
    $\phi$ & nodal electrostatic potential \\
    $E$ & electric field obtained from the discrete potential gradient \\
    $\Gamma_e, \Gamma_i$ & electron and ion face fluxes \\
    $\delta\Gamma_s^\theta$ & learned bounded face flux correction for species $s$ \\
    $\Delta x, \Delta t$ & grid spacing and explicit time step \\
    $Q^k$ & domain integrated discrete charge at step $k$ \\
    $\mathcal{R}_Q$ & one step discrete charge budget residual \\
    $\widetilde{n}$ & raw density before clipping, projection, or flux limiting \\
    \bottomrule
  \end{tabular}
\end{table}

\subsection{Long Horizon Learned PDE Rollout}
A learned one step map can be written abstractly as
\begin{equation}
  \widehat{u}^{k+1} = \mathcal{N}_{\theta}(\widehat{u}^{k}),
\end{equation}
where $u^k$ is a discretized physical state and $\mathcal{N}_{\theta}$ is trained to minimize a local prediction loss. Long horizon prediction applies this map recursively. If the true dynamics preserve a discrete quantity $I(u)$ but the learned map satisfies $I(\mathcal{N}_{\theta}(u)) = I(u) + \epsilon(u)$, then the rollout error in $I$ can accumulate over many steps even when $\epsilon(u)$ is small locally. The same logic applies to density signs: a one step prediction may have small average MSE while producing local negative densities that later alter fields, fluxes, and subsequent predictions.

For this reason, we separate local accuracy metrics from physical fidelity metrics. One step MSE measures interpolation quality. Rollout MSE measures accumulated state error. Charge budget error and minimum density measure whether the predicted trajectory remains physically interpretable. The paper's central claim concerns this long horizon setting, not only the supervised one step problem.

\subsection{Drift Diffusion Poisson Benchmark}
We consider a nondimensional transient one dimensional drift diffusion Poisson system with electron density $n_e(x,t)$, ion density $n_i(x,t)$, electrostatic potential $\phi(x,t)$, and electric field $E(x,t) = -\partial_x \phi(x,t)$. The charged species satisfy
\begin{equation}
  \partial_t n_e + \partial_x \Gamma_e = 0,
  \qquad
  \partial_t n_i + \partial_x \Gamma_i = 0,
\end{equation}
with Drift Diffusion fluxes
\begin{equation}
  \Gamma_e = -\mu_e n_e E - D_e \partial_x n_e + \delta \Gamma_e^\theta,
  \qquad
  \Gamma_i = \mu_i n_i E - D_i \partial_x n_i + \delta \Gamma_i^\theta,
\end{equation}
where $\mu_s$ and $D_s$ are mobility and diffusivity coefficients, and $\delta \Gamma_s^\theta$ is the learned face flux correction. The electrostatic coupling is
\begin{equation}
  -\partial_x^2 \phi = \rho, \qquad \rho = n_i - n_e.
\end{equation}

The present study uses a simplified but explicit boundary treatment. Electrostatic potential uses Dirichlet boundary values, while both charged species use zero wall flux transport boundaries. This is sufficient for controlled experiments on conservation, positivity, and long horizon stability, but it is not yet a physically complete sheath wall model with collection, emission, or Bohm like transport closure. We therefore position the benchmark as a controlled drift diffusion Poisson conservation benchmark rather than a full wall physics simulator.

\subsection{Data Generation}
All reported trajectories are generated by the reference simulator. Unless a sweep changes a field explicitly, the headline nondimensional configuration uses $N_x=16$, domain length $L=1$, $\Delta x=1/16$, $\Delta t=5\times10^{-4}$, $\mu_e=1.0$, $\mu_i=0.4$, $D_e=0.8$, $D_i=0.2$, permittivity $\epsilon=1$, $\phi(0)=0$, $\phi(L)=-3$, zero species wall fluxes, $64$ training trajectories, $64$ supervised steps per trajectory, $8$ warm up steps, and a $20\%$ validation split. Initial densities are generated from sinusoidal perturbations around unit density plus independent Gaussian noise, then clipped to a minimum reference density of $10^{-3}$:
\begin{align}
  n_e(x,0) = \max\!\left(1 + a\sin(2\pi x/L) + \eta_e(x), n_{\min}^{\mathrm{ref}}\right),
  \\
  n_i(x,0) = \max\!\left(1 - a\sin(2\pi x/L) + \eta_i(x), n_{\min}^{\mathrm{ref}}\right),
\end{align}
with default amplitude $a=0.05$, noise scale $0.01$, and trajectory specific random seeds. Source, recombination, ionization, absorbing wall, and emission terms are absent. The resulting trajectory family is deliberately narrow: it probes conservation and rollout stability around smooth near equilibrium density profiles rather than broad plasma operating conditions. This choice keeps the benchmark focused on long horizon conservation and positivity rather than on plasma chemistry or wall modeling.

The target generator uses the same conservative drift diffusion Poisson backbone as the proposed model unless the teacher mismatch configuration is activated. In the mismatch family, the data generator adds a hidden nonlinear face flux term controlled by \texttt{hidden\_flux\_scale}; the learned model is not given this term explicitly. This design is useful for probing whether flux learning can compensate for a missing closure, but it remains a synthetic mismatch rather than an experimentally calibrated plasma model. It also means that the main benchmark partly tests whether a model with the correct numerical inductive bias can recover data generated by a closely related numerical form. Full configuration values for all $64$ included comparisons are listed in Appendix Table~\ref{tab:all_active_evaluations}.

\section{Method}
\subsection{Compatible Discrete Representation}
The discrete state is represented on topology compatible objects: species densities live in cells, transport fluxes live on faces, and electrostatic potential lives on nodes. This assignment is simple, but it matters. With nodal potential, the electric field is recovered by a fixed discrete gradient rather than predicted independently, so field potential compatibility is enforced by construction rather than encouraged by a loss term.

If $\phi_j$ denotes the nodal potential and $\Delta x$ the cell width, then the cell centered electric field is
\begin{equation}
  \bar{E}_{j} = -\frac{\phi_{j+1} - \phi_j}{\Delta x}.
\end{equation}
The face centered field used in the transport fluxes is then obtained by averaging adjacent cell centered values at interior faces,
\begin{equation}
  E_{j+\frac{1}{2}} = \frac{1}{2}\left(\bar{E}_{j} + \bar{E}_{j+1}\right),
\end{equation}
with one sided copying at the domain boundaries. On a uniform grid this two step averaging is algebraically equivalent to the standard second order centered difference $E_{j+\frac{1}{2}} = (\phi_j - \phi_{j+2})/(2\Delta x)$ at node $j{+}1$; it is used here for consistency with the cellwise network inputs. After each transport update, Poisson's equation is solved on the same one dimensional grid with the imposed Dirichlet potential boundaries. Thus the electrostatic state is always recomputed from the updated charge density rather than advanced independently.

The implemented face flux uses centered density interpolation and centered density gradients on interior faces:
\begin{equation}
  \Gamma_{s,j+\frac{1}{2}}^{(0)}
  =
  q_s \mu_s
  \left(\frac{n_{s,j}+n_{s,j+1}}{2}\right)
  E_{j+\frac{1}{2}}
  -
  D_s\frac{n_{s,j+1}-n_{s,j}}{\Delta x},
  \qquad q_e=-1,\;q_i=1.
\end{equation}
Boundary species fluxes are set to zero in the present benchmark. The Poisson solve uses the tridiagonal second difference operator on interior potential nodes with Dirichlet boundary values. For interior node $j$, the right hand side uses the adjacent cell average
\begin{equation}
  \rho_j^{\mathrm{node}} = \frac{1}{2}\left(\rho_{j-1}^{\mathrm{cell}}+\rho_j^{\mathrm{cell}}\right).
\end{equation}
This cell to node averaging is a second order approximation; it does not satisfy an exact discrete charge compatibility identity, so the resolved potential is consistent with the cell charge to second order in $\Delta x$ rather than exactly. In practice, the Poisson solve is always recomputed from the updated densities after every transport step, so electrostatic and density states remain coupled throughout rollout. Time advancement is explicit Euler. No Scharfetter Gummel exponential fitting, adaptive time stepping, source terms, or nonzero wall flux closure is implemented in the current experiments. The centered drift flux is adequate for this controlled nondimensional benchmark, but it is not the preferred final discretization for strongly drift dominated plasma transport. Scharfetter Gummel or upwind compatible variants are natural next baselines.

\subsection{Conservative Update}
Let $n_{s,j}^k$ denote the cell averaged density of species $s$ in cell $j$ at time level $k$, and let $\Gamma_{s,j+\frac{1}{2}}^k$ denote the face flux between adjacent cells. The update takes the finite volume form
\begin{equation}
  n_{s,j}^{k+1}
  =
  n_{s,j}^{k}
  -
  \frac{\Delta t}{\Delta x}
  \left(
    \Gamma_{s,j+\frac{1}{2}}^{k}
    -
    \Gamma_{s,j-\frac{1}{2}}^{k}
  \right).
\end{equation}
Because adjacent cells share the same interior face flux, the interior contributions cancel exactly in the domain integrated budget. Under zero species wall flux boundaries, each species mass and therefore the net charge change only through numerical roundoff, apart from the final finite precision safeguard described below. This algebraic cancellation is the structural property that the direct baseline lacks; note that post hoc global charge projection can partially recover the charge accounting metric (Table~\ref{tab:intermediate_baselines}), but does not reproduce local finite volume flux structure or the associated rollout stability.

We monitor the discrete charge budget through
\begin{equation}
  \mathcal{R}_Q
  =
  Q^{k+1} - Q^{k} + \Delta t\,F_{\partial \Omega},
  \qquad
  Q^{k} = \sum_j \left(n_{i,j}^k - n_{e,j}^k\right)\Delta x,
\end{equation}
where $F_{\partial \Omega}$ is the net boundary charge flux. For the present zero wall flux benchmark, $F_{\partial \Omega}=0$.

\subsection{Positivity Aware Flux Handling}
Simple post update clipping destroys conservation by altering cell masses after the flux balance has already been computed. Instead, the conservative solver limits outgoing fluxes before the update when a proposed step would drive a cell negative. Let $\widetilde{n}_{s,j}^{k+1}$ denote the raw before limiter update. For each cell we define a limiter coefficient
\begin{equation}
  \theta_{s,j}
  =
  \min\!\left(
    1,\;
    \frac{n_{s,j}^{k}}{\mathcal{O}_{s,j}^{k} + \varepsilon}
  \right),
\end{equation}
where $\mathcal{O}_{s,j}^{k}$ is the outgoing mass implied by the proposed face fluxes and $\varepsilon > 0$ is a small numerical safeguard (set to $10^{-8}$ in all experiments, chosen to be well below the initial density scale of order unity). Outgoing face fluxes are scaled by the neighboring $\theta_{s,j}$ values before the conservative update is applied. This design preserves the discrete budget while reducing the risk of nonphysical densities. Benchmark results show that this mechanism is highly effective, though not inactive in every stress setting; the grid-64 case activates the floor correction path even though the final clamp mass remains zero.

More explicitly, for a cell $j$ the implementation computes
\begin{equation}
  \mathcal{O}_{s,j}^{k}
  =
  \frac{\Delta t}{\Delta x}
  \left[
    \max(\Gamma_{s,j+\frac{1}{2}}^k,0)
    +
    \max(-\Gamma_{s,j-\frac{1}{2}}^k,0)
  \right],
\end{equation}
then scales only flux components that remove mass from the donor cell. If $\Gamma_{s,j+\frac{1}{2}}>0$, the face flux is multiplied by $\theta_{s,j}$; if $\Gamma_{s,j+\frac{1}{2}}<0$, it is multiplied by $\theta_{s,j+1}$. The learned correction is limited together with the physical Drift Diffusion flux, and each face retains a single shared value after limiting, so the antisymmetry needed for finite volume cancellation is preserved. The limiter is used during both training and rollout, and gradients are propagated through its tensor implementation.

A final nonnegative redistribution/clamp is present in the implementation as a numerical safeguard. The redistribution step is designed to remove tiny negative values while returning the removed mass to available positive cells. If all available mass is exhausted, a residual clamp can break exact species mass conservation at finite precision. The evaluation code records the safeguard correction fraction, redistributed deficit, final clamp fraction, final clamp mass, and absolute mass change introduced by this safeguard. Table~\ref{tab:safeguard_diagnostics} reports these diagnostics for representative headline and stress cases. The possibility of a final finite precision clamp prevents us from claiming a formal strict positivity proof.

\subsection{Learned Flux Correction}
The learned component predicts corrections on the transport path instead of predicting the full next state. Operationally, the network receives seven cellwise input channels:
\begin{equation}
  (n_e,\; n_i,\; \phi_{\mathrm{left}},\; \phi_{\mathrm{right}},\; E,\; n_i-n_e,\; x/L).
\end{equation}
A shared cellwise multilayer perceptron with Tanh activations maps these features to hidden states. Face hidden states are formed by averaging neighboring cell hidden states, with one sided copying at boundary faces, and a linear decoder returns two face corrections, one for each species. The default headline model uses hidden size $64$, three layers, and correction scale $0.05$, so the actual correction is $0.05\tanh(\delta\Gamma^\theta)$. The same translationally shared network is used across faces, while separate output channels represent electron and ion corrections. For the headline grid, \method\ has $4{,}802$ trainable parameters. The model is local and translationally shared, but it is not a resolution invariant neural operator; the current grid resolution studies retrain at each resolution.

This keeps the learned component small and interpretable: the neural model is asked to correct transport behavior, while the backbone retains control over conservation bookkeeping. The ablation results below show that the structural core is the dominant stabilizing ingredient in the present benchmark; the learned correction should therefore be read as an extensible closure mechanism rather than as the sole source of the reported stability.

\subsection{Training Objectives}
Training uses supervised next step targets together with physically motivated monitoring terms. For both models, the optimization objective is
\begin{equation}
  \mathcal{L}
  =
  \mathcal{L}_{\mathrm{sup}}
  +
  \lambda_{\mathrm{pos}}\,\mathcal{L}_{\mathrm{pos}}
  +
  \lambda_{\mathrm{cons}}\,\mathcal{L}_{\mathrm{cons}},
\end{equation}
where $\mathcal{L}_{\mathrm{sup}}$ is the one step supervised mean squared error and $\mathcal{L}_{\mathrm{pos}}$ is
\begin{equation}
  \mathcal{L}_{\mathrm{pos}}
  =
  \mathbb{E}_{s\in\{e,i\},j}\left[\mathrm{ReLU}(-\widetilde{n}_{s,j})^2\right],
\end{equation}
evaluated on the raw before projection densities. The term $\mathcal{L}_{\mathrm{cons}} = \mathbb{E}[\mathcal{R}_Q^2]$ monitors discrete charge balance error. The important design choice is that conservation is not delegated solely to a soft penalty. Instead, the update itself is conservative, and the training objective is used to shape accuracy, regularity, and positivity behavior around that exact core. The implementation records loss histories, raw negativity diagnostics, conservation related diagnostics, rollout charge error, and minimum density behavior so that model selection can be based on physically relevant outcomes rather than one step fit alone.

Both main models are trained with Adam using the configuration specified learning rate, default $3\times 10^{-4}$, batch size $32$, no weight decay, and the best validation checkpoint selected by total validation loss. The headline run uses $100$ epochs, $\lambda_{\mathrm{pos}}=0.1$, and $\lambda_{\mathrm{cons}}=0.1$. The primary objective is one step supervised training; scheduled sampling is not used. We additionally include a four step rollout trained \baseline\ diagnostic in the headline comparison. The same conservation term therefore has asymmetric meaning across models. For \method, $\mathcal{L}_{\mathrm{cons}}$ is usually near zero because the update is already conservative, so this term mainly acts as a diagnostic and safeguard. For \baseline, the conservation term is a soft penalty computed from the change in predicted domain integrated charge between the input and predicted next state.

\subsection{Baseline Comparator}
The baseline, \baseline, directly regresses a residual update from the flattened current features to the next cellwise $(n_e,n_i,\phi_{\mathrm{cc}})$ state. It uses a fully connected Tanh network with the same default hidden size and depth setting as the conservative model; for the headline grid it has $14{,}512$ trainable parameters. Predicted electron and ion densities are clamped to be nonnegative before evaluation, while potential is predicted directly and converted to nodal values for electric field reconstruction. Thus repeated rollout minimum densities near $10^{-14}$ should be read as numerical near zero behavior after direct prediction and nonnegative clamping, not as evidence of a physically meaningful density floor.

We additionally evaluate three direct model variants. The first recomputes $\phi$ exactly from the predicted densities by solving Poisson after each direct density prediction. The second applies a diagnostic global charge projection after each direct prediction by adding the required uniform nonnegative correction to one species so that $\sum_j(n_{i,j}-n_{e,j})$ matches the input charge. This projection is intentionally simple and should not be read as the strongest possible projected state model. The third trains the same direct architecture with a four step autoregressive rollout loss. These variants do not introduce local shared face flux structure; they test whether Poisson compatibility, global charge correction, or short rollout training alone explains the conservative model's advantage.

\section{Experimental Protocol}
\subsection{Experiment Stack}
The experiment suite follows an ordered protocol comprising headline training, multiseed repeats, harder regime stress tests, core ablations, rollout stability sweeps, teacher mismatch tests, parameter generalization sweeps, mesh resolution studies, data regime studies, training duration studies, model capacity studies, conservation weight studies, and positivity activation studies. The benchmark comprises $64$ prespecified configurations: the single headline experiment together with the multirun families listed in Appendix Table~\ref{tab:all_active_evaluations}. Exploratory runs conducted before the benchmark protocol was finalized are excluded; only configurations belonging to the formalized protocol directories are counted and used to support the paper's claims.

The headline configuration uses $64$ trajectories and $64$ supervised steps, yielding $4{,}096$ supervised training examples, followed by $100$ training epochs and evaluation rollouts of up to $128$ steps over $16$ evaluation trajectories. Additional stages systematically vary horizon length, data quantity, transport stiffness, physical parameters, mesh resolution, architectural capacity, correction scale, and loss weights. Stage names are descriptive labels for configuration changes: for example, ``data sparse'' uses $8$ trajectories, ``stiff transport'' changes the transport coefficients and time step, and ``long horizon sparse'' combines sparse data with a $256$-step evaluation rollout.

\subsection{Metrics}
We report four primary metrics: one step mean squared error (MSE), rollout MSE over multistep prediction, rollout charge error from the discrete domain integrated charge budget, and rollout minimum density as a practical admissibility diagnostic. The first metric measures local prediction fidelity. The remaining three target what actually matters for reliable scientific rollout: accumulated state error, conservative charge accounting, and density admissibility.

The MSE metrics are unnormalized arithmetic means over the predicted cellwise target channels $(n_e,n_i,\phi_{\mathrm{cc}})$. One step MSE is averaged over supervised samples, cells, and channels. Rollout MSE is evaluated autoregressively against the reference trajectory at every rollout step and averaged over evaluation trajectories, rollout steps, cells, and the same three channels. Electric field is not included directly in the primary MSE; it is reconstructed from potential for rollout state advancement and diagnostics. Table~\ref{tab:headline_channel_errors} reports the headline channel wise decomposition, including an electric field diagnostic, to show which variables dominate the aggregate error.

\begin{table}[t]
  \centering
  \caption{Headline channel wise error decomposition. The first three columns are MSEs for the supervised target channels. The electric field column is a reconstructed rollout diagnostic and is not included in the primary aggregate MSE.}
  \label{tab:headline_channel_errors}
  \resizebox{\linewidth}{!}{
  \begin{tabular}{llcccc}
    \toprule
    Setting & Model & $n_e$ MSE & $n_i$ MSE & $\phi_{\mathrm{cc}}$ MSE & $E$ MSE \\
    \midrule
    One step & \method & $4.17\times 10^{-12}$ & $4.02\times 10^{-13}$ & $3.16\times 10^{-14}$ & n/a \\
    One step & \baseline & $3.91\times 10^{-7}$ & $2.64\times 10^{-7}$ & $2.08\times 10^{-7}$ & n/a \\
    Rollout & \method & $2.05\times 10^{-8}$ & $1.43\times 10^{-9}$ & $1.21\times 10^{-10}$ & $1.29\times 10^{-9}$ \\
    Rollout & \baseline & $8.38\times 10^{1}$ & $4.07\times 10^{1}$ & $2.40\times 10^{0}$ & $9.10\times 10^{1}$ \\
    \bottomrule
  \end{tabular}
  }
\end{table}

The reported rollout charge error is the mean squared absolute charge budget residual over all rollout steps and evaluation trajectories:
\begin{equation}
  \mathrm{CE}_{\mathrm{rollout}}
  =
  \mathbb{E}_{r,k}
  \left[
    \left(
      Q_{r,k}^{\mathrm{pred}} - Q_{r,0}^{\mathrm{pred}}
      + \sum_{\ell=0}^{k-1}\Delta t\,F_{\partial\Omega,r,\ell}^{\mathrm{pred}}
    \right)^2
  \right].
\end{equation}
It is absolute rather than relative. In the present zero wall flux benchmark, $F_{\partial\Omega}=0$ for the reference physics, while the conservative evaluator still records the explicit boundary flux term from the predicted face fluxes. Rollout minimum density is the minimum post update density over species, cells, rollout steps, and evaluation trajectories.

For the positivity activation family we additionally report raw negativity diagnostics computed before positivity projection: the raw negative fraction
\begin{equation}
  f_{\mathrm{neg}}
  =
  \frac{1}{2N}
  \sum_{j,s} \mathbf{1}\!\left[\widetilde{n}_{s,j}<0\right],
\end{equation}
the mean and maximum negative magnitudes, and the raw minimum density
\begin{equation}
  n_{\min}^{\mathrm{raw}} = \min_{j,s} \widetilde{n}_{s,j}.
\end{equation}
These diagnostics are not used as headline metrics for the entire paper; they are used specifically to interpret whether the positivity stress benchmark truly activates unconstrained negative proposals.

\subsection{Interpretation Rule}
When short horizon error and physical fidelity disagree, we report the disagreement rather than collapse it into a single winner. For this benchmark, we treat large charge budget drift or density collapse as disqualifying for physically interpretable plasma rollout. In applications where small absolute charge drift is acceptable, a lower rollout MSE model may be preferable; the tradeoff cases below should therefore be interpreted through the user's tolerance for conservation error, not as a universal ranking rule.

\section{Main Results}
\subsection{Headline Comparison}
\begin{figure}[t]
  \centering
  \begin{subfigure}[t]{0.48\linewidth}
    \centering
    \includegraphics[width=\linewidth]{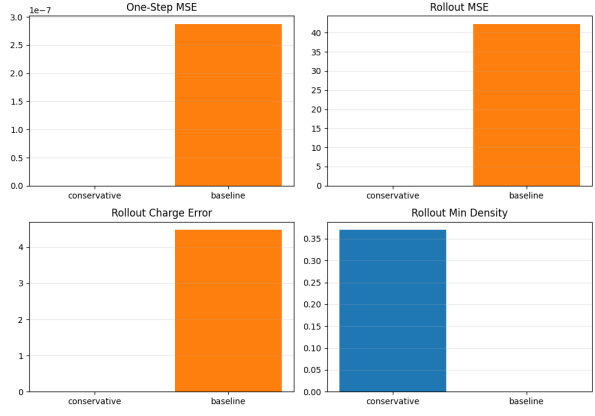}
    \caption{Headline metric summary.}
  \end{subfigure}
  \hfill
  \begin{subfigure}[t]{0.48\linewidth}
    \centering
    \includegraphics[width=\linewidth]{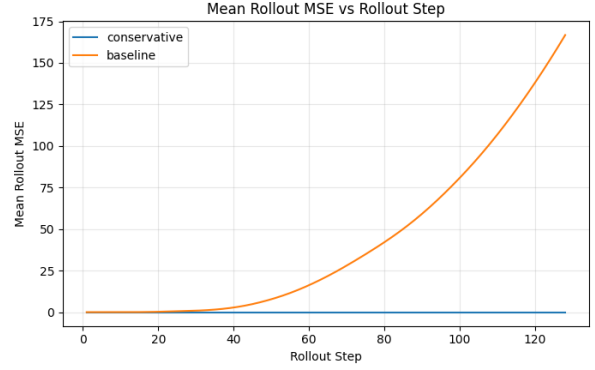}
    \caption{Mean rollout MSE over time.}
  \end{subfigure}

  \vspace{0.4em}

  \begin{subfigure}[t]{0.48\linewidth}
    \centering
    \includegraphics[width=\linewidth]{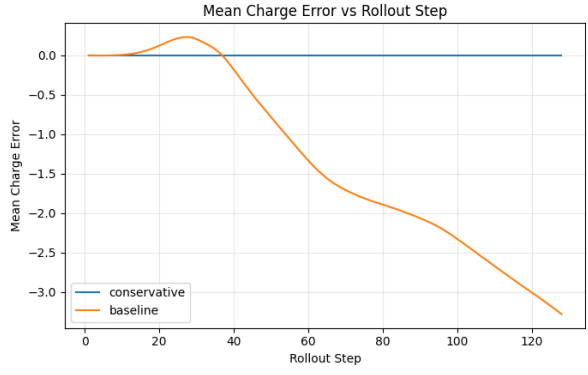}
    \caption{Rollout charge error history.}
  \end{subfigure}
  \hfill
  \begin{subfigure}[t]{0.48\linewidth}
    \centering
    \includegraphics[width=\linewidth]{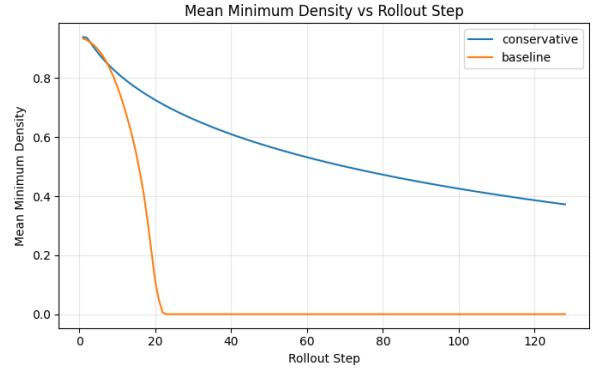}
    \caption{Rollout minimum density history.}
  \end{subfigure}
  \caption{Headline experiment results. MSE is averaged over $(n_e,n_i,\phi_{\mathrm{cc}})$, cells, rollout steps, and evaluation trajectories; charge error is the mean squared absolute charge budget residual. The conservative solver preserves charge accounting by algebraic construction and remains stable over rollout, whereas the direct baseline degrades rapidly.}
  \label{fig:headline_metrics}
\end{figure}

Figure~\ref{fig:headline_metrics} and Table~\ref{tab:headline} summarize the main comparison. The conservative solver and the direct baseline do not occupy the same regime of behavior in this benchmark. On the headline run, \method\ reaches a rollout charge error metric of $5.93\times 10^{-15}$, consistent with roundoff level conservation for the stated nondimensional scale, while the baseline accumulates $4.48$. This charge result is guaranteed by the shared face zero wall flux update. It is an enforced structural property, not an empirical finding, so the relevant empirical question is the associated rollout accuracy and profile behavior. Table~\ref{tab:headline} also shows the classical core only result, which is more accurate than the learned variant on rollout MSE ($1.15\times 10^{-14}$ versus $7.35\times 10^{-9}$), confirming that the conservative structure, not the learned correction, is the dominant stabilizing factor. The same run shows rollout MSE $7.35\times 10^{-9}$ for the conservative solver versus $4.23\times 10^{1}$ for the baseline. The minimum density also stays positive for the conservative solver, whereas the direct baseline approaches numerical zero.

\begin{table}[t]
  \centering
  \caption{Headline experiment. Lower is better for one step MSE, rollout MSE, and rollout charge error; higher is better for rollout minimum density. The row for the classical core without learned correction sets the correction scale to zero, keeping the conservative update, Poisson reconstruction, and positivity aware transport; it establishes that conservative discrete structure is the dominant factor, not the learned correction. Charge error near roundoff is enforced by the shared face update under zero wall fluxes, not learned.}
  \label{tab:headline}
  \resizebox{\linewidth}{!}{
  \begin{tabular}{lcccc}
    \toprule
    Model & One step MSE & Rollout MSE & Rollout charge error & Rollout min density \\
    \midrule
    \method & $1.54\times 10^{-12}$ & $7.35\times 10^{-9}$ & $5.93\times 10^{-15}$ & $3.70\times 10^{-1}$ \\
    Classical core only (no learned correction) & $1.67\times 10^{-14}$ & $1.15\times 10^{-14}$ & $6.88\times 10^{-15}$ & $3.70\times 10^{-1}$ \\
    \baseline & $2.88\times 10^{-7}$ & $4.23\times 10^{1}$ & $4.48\times 10^{0}$ & $1.00\times 10^{-14}$ \\
    \bottomrule
  \end{tabular}
  }
\end{table}

\begin{table}[t]
  \centering
  \caption{Intermediate direct baseline checks. Poisson recomputation improves the direct baseline but leaves large rollout error and charge drift. Global charge projection fixes the charge metric but worsens rollout MSE, because the added uniform density offset alters the local charge distribution, changes the Poisson solve, and feeds back into transport; this projection is a diagnostic check, not the strongest possible projected state model. Four step rollout training improves short horizon behavior but still leaves large long horizon error. Rollout minimum density is $\leq 1.00\times 10^{-14}$ for all \baseline\ variants, reflecting nonnegative clamping to numerical near zero.}
  \label{tab:intermediate_baselines}
  \resizebox{\linewidth}{!}{
  \begin{tabular}{lcccc}
    \toprule
    Model & One step MSE & Rollout MSE & Rollout charge error & Rollout min density \\
    \midrule
    \baseline & $2.88\times 10^{-7}$ & $4.23\times 10^{1}$ & $4.48\times 10^{0}$ & $1.00\times 10^{-14}$ \\
    \baseline\ + Poisson recomputation & $2.18\times 10^{-7}$ & $2.53\times 10^{1}$ & $2.60\times 10^{-1}$ & $1.00\times 10^{-14}$ \\
    \baseline\ + charge projection & $2.87\times 10^{-7}$ & $6.72\times 10^{1}$ & $1.34\times 10^{-12}$ & $1.00\times 10^{-14}$ \\
    \baseline\ + rollout training & $4.16\times 10^{-7}$ & $2.71\times 10^{1}$ & $2.77\times 10^{0}$ & $1.00\times 10^{-14}$ \\
    \bottomrule
  \end{tabular}
  }
\end{table}

\begin{figure}[!h]
  \centering
  \includegraphics[width=0.82\linewidth]{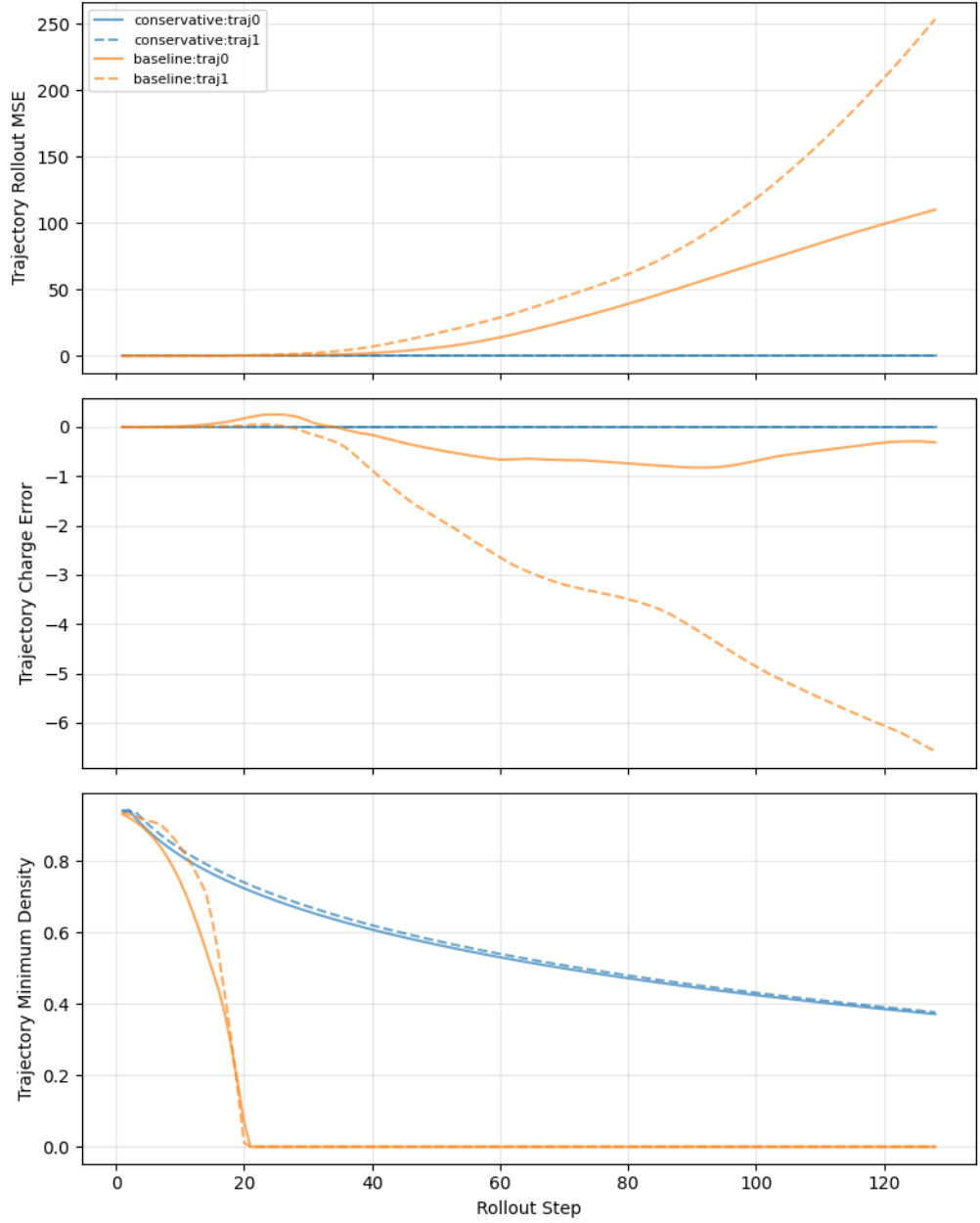}
  \caption{Representative headline rollout trajectories for two evaluation initial conditions. The panel shows trajectory level rollout MSE, charge error, and minimum density over time. This metric level view diagnoses when the baseline rollout begins to leave the reference dynamics.}
  \label{fig:trajectory_examples}
\end{figure}

\begin{figure}[!h]
  \centering
  \includegraphics[width=\linewidth]{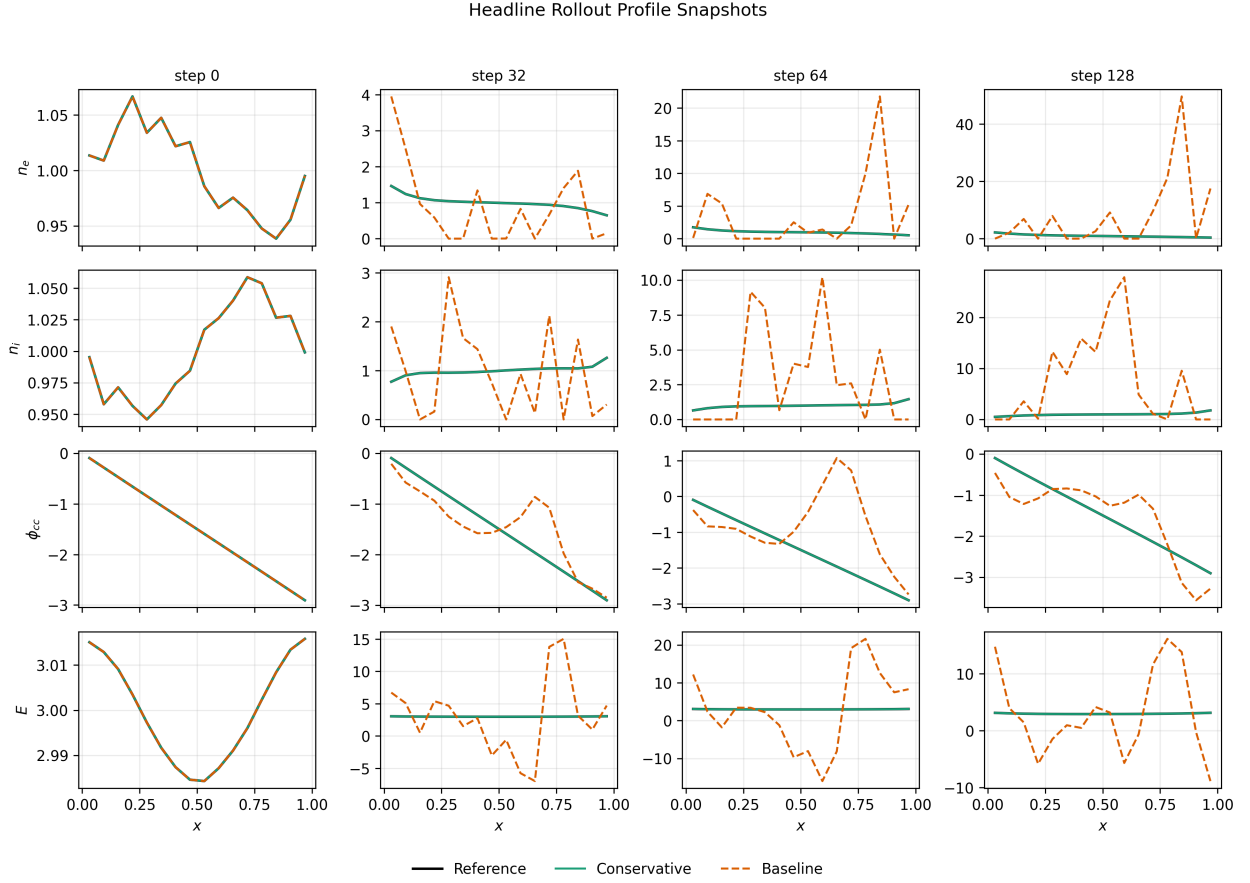}
  \caption{Headline physical profile snapshots for one evaluation trajectory at rollout steps $0$, $32$, $64$, and $128$. The rows show electron density, ion density, cell centered potential, and reconstructed electric field. The conservative rollout remains close to the reference profile, while the direct baseline develops large density and field distortions during autoregressive rollout.}
  \label{fig:profile_snapshots}
\end{figure}

Figure~\ref{fig:profile_snapshots} shows the corresponding physical profiles for one headline trajectory. The conservative and reference profiles remain visually close through the rollout. The direct baseline develops large density excursions and electric field distortions, which explains why its aggregate rollout MSE and charge budget error grow together. This profile view is still only a diagnostic for the controlled benchmark; it is not evidence of realistic sheath wall behavior.

Table~\ref{tab:intermediate_baselines} separates possible explanations for the direct baseline failure. Exact Poisson recomputation reduces rollout MSE from $42.3$ to $25.3$ and charge error from $4.48$ to $0.260$, showing that electrostatic incompatibility is part of the failure. It does not close the gap to the conservative model. Global charge projection reduces the charge error metric to $1.34\times 10^{-12}$, but rollout MSE worsens to $67.2$. A four step rollout trained direct model gives rollout MSE $27.1$ and charge error $2.77$, so short rollout training alone also does not reproduce the effect of local shared face transport. Thus neither global charge correction, exact electrostatic recomputation, nor short autoregressive training is equivalent to local finite volume structure in this benchmark.

\subsection{Seed Robustness}
The multiseed stage confirms that the headline result is not a single seed artifact. Across five independent seeds, the conservative solver maintains mean rollout charge error $6.76\times 10^{-15} \pm 1.79\times 10^{-15}$ (near roundoff by construction) and mean rollout MSE $7.76\times 10^{-9} \pm 2.10\times 10^{-9}$. The baseline, by contrast, shows mean rollout charge error $1.36 \pm 0.98$ and mean rollout MSE $4.44\times 10^{1} \pm 2.95\times 10^{1}$. The minimum density also remains stable for the conservative solver and approaches numerical zero for the baseline in every seed. Exact aggregate values are listed in Appendix Table~\ref{tab:multiseed}.

\subsection{Harder Regime Stress Tests}
The conservative advantage persists in more difficult settings, including stronger bias, sparse data, stiff transport, and long horizon sparse training. In the most severe long horizon sparse setting, the conservative solver holds rollout charge error to $1.60\times 10^{-14}$ (near roundoff, by construction), while the baseline rises to $1.13\times 10^{2}$. Even when the conservative rollout MSE grows under stress, it remains substantially below the baseline across the benchmark suite. Representative exact values are collected in Appendix Table~\ref{tab:stress}.

\subsection{Cross Stage Summary}
\begin{figure}[t]
  \centering
  \includegraphics[width=\linewidth]{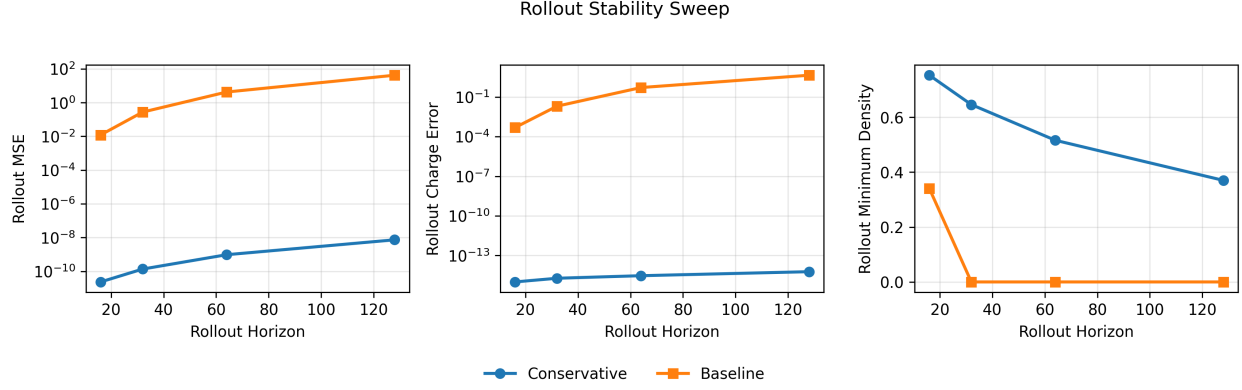}
  \caption{Rollout stability sweep across horizons 16, 32, 64, and 128. The conservative solver retains roundoff level charge error metrics while rollout MSE and minimum density behavior remain more stable than the direct baseline. Curves show single configuration means rather than multiseed uncertainty bands.}
  \label{fig:stability_sweep}
\end{figure}

Taken together, the benchmark shows that the conservative solver achieves near roundoff charge error by construction in all $64$ configurations, wins rollout MSE in $60/64$, and wins rollout minimum density in $61/64$, while winning one step MSE in only $19/64$. The full registry of benchmark configurations is provided in Appendix Table~\ref{tab:all_active_evaluations}. This asymmetry supports the paper's main point within the tested comparison class: local prediction accuracy and long horizon physical fidelity are related but not equivalent objectives. Figure~\ref{fig:stability_sweep} shows that the rollout MSE separation widens steadily with rollout horizon.

\section{Physical Fidelity Versus Short Horizon Accuracy}
\begin{figure}[t]
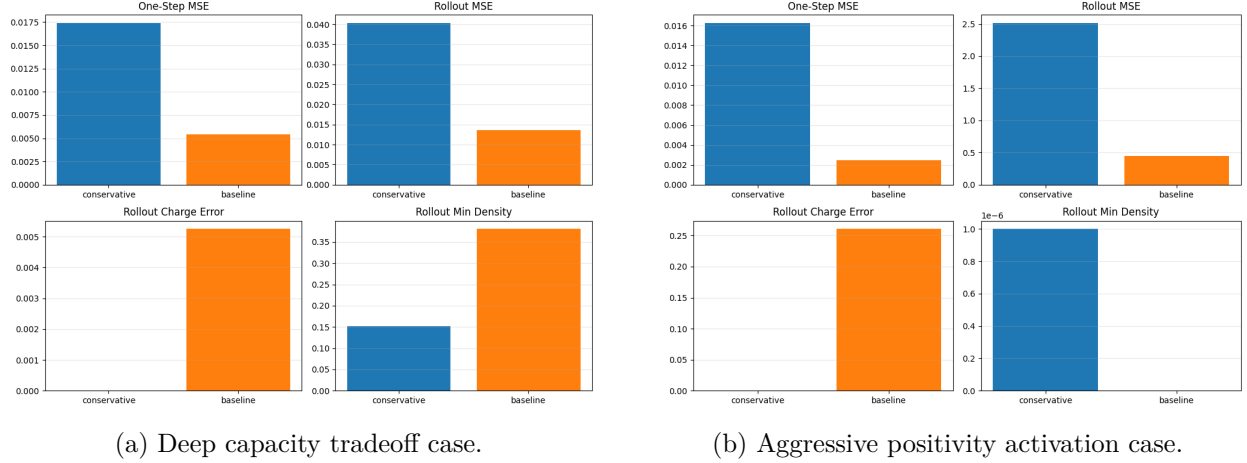

  \centering
  \begin{subfigure}[t]{0.48\linewidth}
    \centering
    \includegraphics[width=\linewidth]{\manufig{layers4_tradeoff_bars.png}}
    \caption{Deep capacity tradeoff case.}
  \end{subfigure}
  \hfill
  \begin{subfigure}[t]{0.48\linewidth}
    \centering
    \includegraphics[width=\linewidth]{\manufig{positivity1_tradeoff_bars.png}}
    \caption{Aggressive positivity activation case.}
  \end{subfigure}
  \caption{Tradeoff cases in which short horizon or rollout MSE and charge accounting favor different model choices. In both cases the conservative solver keeps charge accounting near roundoff level while the baseline incurs larger charge drift.}
  \label{fig:tradeoff_cases}
\end{figure}

The benchmark results contain important counterexamples to MSE only evaluation. In the deepest model capacity setting, the baseline outperforms the conservative solver on one step MSE and rollout MSE, and it also achieves a larger minimum density. However, the baseline rollout charge error metric is still $5.25\times 10^{-3}$ versus $1.08\times 10^{-14}$ for the conservative solver. Whether this difference is acceptable depends on the target application's conservation tolerance. Our interpretation is that the conservative model is preferable when charge accounting is a hard requirement, while the direct baseline can be preferable when the sole objective is unnormalized rollout MSE. Figure~\ref{fig:tradeoff_cases} makes this mismatch explicit.

\begin{figure}[t]
  \centering
  \includegraphics[width=\linewidth]{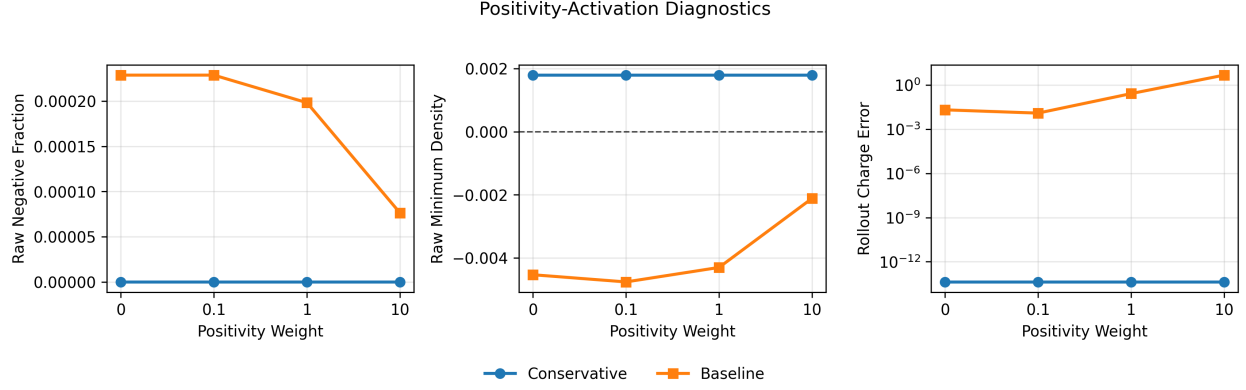}
  \caption{Training raw negativity diagnostics across positivity weights $0$, $0.1$, $1$, and $10$. The conservative solver remains nearly flat because its conservative transport path and flux limiter already suppress raw negativity during training, whereas the baseline reduces raw negatives only partially and does so without preserving the same charge budget quality.}
  \label{fig:positivity_activation}
\end{figure}

The positivity activation stress stage makes a complementary diagnostic point. This family is not an accuracy benchmark; it was constructed to force raw negative updates near the density floor. In that setting, increasing the baseline positivity weight reduces the maximum training raw negative fraction from $1.94\times 10^{-1}$ to $2.54\times 10^{-2}$ and improves the worst raw minimum density from $-3.79\times 10^{-1}$ to $-4.08\times 10^{-2}$. By contrast, the conservative solver remains effectively insensitive to the positivity weight sweep during training, which indicates that its admissibility behavior is governed primarily by the conservative transport structure and flux limiter rather than the positivity loss coefficient. Figure~\ref{fig:positivity_activation} shows that the conservative model is already structurally protected against raw negativity, whereas the baseline trades off raw negativity regularization against weaker charge accounting. These cases show why a benchmark centered only on mean squared error would mischaracterize the scientific value of the method.
Detailed side by side numbers for the tradeoff cases are reported in Appendix Table~\ref{tab:tradeoffs}.

\begin{table}[t]
  \centering
  \caption{Conservative model safeguard diagnostics for representative stress cases. Raw minimum density is measured before the positivity limiter and final safeguard during rollout. Rollout minimum density is the post safeguard minimum. The floor correction fraction is the maximum over species, rollout steps, and trajectories; final clamp mass is the maximum mass inserted by the residual clamp after redistribution.}
  \label{tab:safeguard_diagnostics}
  \resizebox{\linewidth}{!}{
  \begin{tabular}{lccccc}
    \toprule
    Case & Raw min density & Rollout min density & Max floor correction fraction & Max redistributed deficit & Max final clamp mass \\
    \midrule
    Headline & $3.70\times 10^{-1}$ & $3.70\times 10^{-1}$ & $0$ & $0$ & $0$ \\
    Grid 64 & $-1.97\times 10^{1}$ & $0.00$ & $2.03\times 10^{-1}$ & $3.73\times 10^{-8}$ & $0$ \\
    Positivity 1.0 & $-2.64\times 10^{-4}$ & $1.00\times 10^{-6}$ & $0$ & $0$ & $0$ \\
    Stiff transport & $1.60\times 10^{-1}$ & $1.60\times 10^{-1}$ & $0$ & $0$ & $0$ \\
    Long horizon sparse & $9.16\times 10^{-2}$ & $9.16\times 10^{-2}$ & $0$ & $0$ & $0$ \\
    Teacher mismatch & $1.51\times 10^{-1}$ & $1.51\times 10^{-1}$ & $0$ & $0$ & $0$ \\
    \bottomrule
  \end{tabular}
  }
\end{table}

\section{Conservative Structure Versus Learned Correction Ablation}
The ablation results confirm that the conservative structure is the dominant source of stability, not the learned correction. The conservative variant without learned correction (correction scale set to zero, conservative update and positivity aware transport retained) already achieves one step MSE $1.67\times 10^{-14}$, rollout MSE $1.15\times 10^{-14}$, and rollout charge error $6.88\times 10^{-15}$. In the headline configuration, this classical conservative core outperforms the full learned correction variant on rollout MSE by more than five orders of magnitude. This paper should therefore be read primarily as evidence for conservative discrete structure, not as evidence that the current learned correction has solved a missing plasma closure.

\begin{table}[t]
  \centering
  \caption{Conservative structure versus learned correction ablation. Rows for the classical core set the learned correction scale to zero while keeping the conservative update, Poisson reconstruction, and positivity aware transport path. The core alone outperforms the learned correction variant on rollout MSE in the headline case, showing that the structural core is the dominant stabilizing factor; the learned correction provides an extensible closure mechanism for settings where the backbone physics is incomplete.}
  \label{tab:core_only_main}
  \resizebox{\linewidth}{!}{
  \begin{tabular}{lcccc}
    \toprule
    Case & One step MSE & Rollout MSE & Rollout charge error & Rollout min density \\
    \midrule
    Headline learned correction & $1.54\times 10^{-12}$ & $7.35\times 10^{-9}$ & $5.93\times 10^{-15}$ & $3.70\times 10^{-1}$ \\
    Classical core without learned correction & $1.67\times 10^{-14}$ & $1.15\times 10^{-14}$ & $6.88\times 10^{-15}$ & $3.70\times 10^{-1}$ \\
    Small correction scale & $5.74\times 10^{-14}$ & $2.50\times 10^{-10}$ & $1.11\times 10^{-14}$ & $3.70\times 10^{-1}$ \\
    Teacher mismatch learned correction & $1.74\times 10^{-2}$ & $4.04\times 10^{-2}$ & $1.63\times 10^{-14}$ & $1.51\times 10^{-1}$ \\
    Teacher mismatch core without learned correction & $1.74\times 10^{-2}$ & $4.06\times 10^{-2}$ & $1.09\times 10^{-14}$ & $1.52\times 10^{-1}$ \\
    \bottomrule
  \end{tabular}
  }
\end{table}

The teacher mismatch family supports a similarly cautious interpretation. In the strengthened teacher mismatch benchmark, the conservative model reports one step MSE $1.74\times 10^{-2}$, rollout MSE $4.04\times 10^{-2}$, rollout charge error $1.63\times 10^{-14}$, and minimum density $1.51\times 10^{-1}$, while the baseline records rollout MSE $1.56$ and rollout charge error $7.70\times 10^{-1}$. However, the corresponding teacher flux variant without learned correction achieves rollout MSE $4.06\times 10^{-2}$, nearly identical to the learned version. The current evidence therefore does not establish that the learned correction is recovering the hidden flux term. A more defensible interpretation is that learning is compatible with the conservative framework and provides a mechanism for future closure settings, but the dominant stabilizing effect comes from the conservative core itself. A convincing learned value result would require a setting where the classical core clearly fails and the learned correction reduces rollout error by at least an order of magnitude, for example through a hidden nonlinear mobility, a Scharfetter Gummel teacher paired with a centered flux backbone, or a hidden wall flux term.

This ablation is central to the paper's scope. The present evidence separates two conclusions. First, finite volume conservation, compatible Poisson reconstruction, and positivity aware limiting are sufficient to explain most of the observed long horizon stability on the baseline benchmark. Second, the learned flux pathway provides a mechanism for future missing closure experiments, but the current synthetic mismatch tests are not yet strong enough to establish a broad learned plasma solver claim. The present draft also does not yet report learned correction magnitude distributions or direct comparisons between the learned correction and the hidden teacher flux. Stronger evidence would require nonzero wall flux physics, noisy targets, data generated by a solver that is less closely matched to the conservative backbone, and stronger projected or neural finite volume comparators.

\section{Generalization and Scaling Behavior}
The method remains strong under parameter shifts, data scarcity, and shortened training schedules. In the stronger bias generalization case, the conservative solver reaches rollout MSE $7.13\times 10^{-2}$ and rollout charge error $1.44\times 10^{-14}$, while the baseline yields $3.72$ and $2.39$, respectively. With only four training trajectories, the conservative solver attains rollout MSE $4.04\times 10^{-2}$ and charge error $2.09\times 10^{-14}$, versus $8.76$ and $5.61\times 10^{-1}$ for the baseline. Even after only two training epochs, the conservative solver still records charge error near $10^{-14}$ and rollout MSE below $0.04$, while the baseline remains near $9.91$ rollout MSE.

The mesh resolution study is more nuanced and requires honest discussion. These are retrained grid resolution cases rather than zero shot resolution transfer tests. At grid size $64$, the conservative solver preserves near zero charge drift ($9.04\times 10^{-15}$, by construction) and still strongly outperforms the baseline on rollout MSE ($0.704$ versus $10.256$), but the raw minimum density before limiting reaches $-1.97\times 10^{1}$. For a density field initialized near unity, this indicates that the centered explicit transport scheme is genuinely unstable at this resolution under the tested settings. It is a CFL type violation of the monotonicity condition, not a small finite precision artifact. The flux limiter and redistribution safeguard recover a nonnegative post update state in this run (maximum floor correction fraction $0.203$, final clamp mass zero), but this is an implemented correction mechanism rather than a stability guarantee. It demonstrates that upwind or Scharfetter Gummel discretizations are likely necessary for the method to remain robust at higher spatial resolution, and not merely optional extensions. Likewise, the positivity activation family should not be read as an accuracy contest; its role is to diagnose whether unconstrained negative proposals are present and whether they can be regularized. Additional exact values are collected in Appendix Table~\ref{tab:generalization}.

\section{Discussion}
\paragraph{Drift Diffusion Poisson models in plasma physics.}
Drift Diffusion Poisson fluid models are a well established reduced description for low temperature plasma transport and discharge modeling \citep{lieberman2005principles,hagelaar2005boltzmann}. They appear in standard plasma physics textbooks and have been applied to glow discharges, radio frequency plasmas, plasma processing, and sheath analysis for several decades. The conservative finite volume discretization tradition for these equations also has deep roots in plasma modeling, tracing to exponentially fitted schemes developed for semiconductor plasma analogies \citep{scharfetter1969large,bessemoulin2012finite}. The present work is therefore not introducing DDP models to plasma physics; rather, it asks whether a learned neural surrogate for DDP dynamics can be constructed in a way that respects the same conservation properties that make classical DDP solvers reliable. This framing positions the contribution as a design study for structure preserving learned surrogates, with DDP as the target system.

\paragraph{Mechanism and structural dominance.}
The study points to a mechanism rather than only a performance ranking. In the direct baseline, the neural network represents the full time advancement map. Any local inconsistency in charge, density sign, or electrostatic coupling becomes part of the next input state, where it can change the Poisson solve, alter drift fluxes, and feed back into subsequent transport. This is the central reason plasma transport is a difficult surrogate learning target: density errors and field errors are not independent output channels, but coupled variables in a closed transport loop. A direct model can therefore remain accurate in a supervised one step sense while drifting into a trajectory that no longer corresponds to the discrete plasma model.

The conservative solver changes this failure mode by restricting what the learned component is allowed to do. Shared face fluxes force interior flux cancellation before the optimizer is considered. Nodal potential and discrete differentiation prevent the model from inventing an electric field that is incompatible with the potential. Positivity aware flux handling acts before the update, so admissibility protection does not require nonconservative post hoc clipping. These mechanisms do not make the model exact; the learned flux correction can still be inaccurate, and the high resolution stress case shows that small positivity violations can remain. They do, however, move the dominant errors away from budget breaking update algebra and toward ordinary approximation error inside a physically constrained map.

This interpretation also clarifies the role of the learned correction. The ablation results do not support the claim that neural flux corrections are the primary source of the observed improvement. The more defensible conclusion is that learning becomes useful only after the discrete structure has been made reliable. In the present benchmark, the classical conservative core already explains most of the stability (and in the headline case outperforms the learned variant on rollout MSE), while the learned component provides a flexible path for future closure modeling and teacher mismatch settings. This is an important distinction for research positioning: the paper is not a generic demonstration that a neural network can fit plasma trajectories, but an argument that learned plasma solvers should be built around the algebra of the governing discretization, with the neural correction as an extensible but currently secondary ingredient.

The results further suggest that learned PDE benchmarks should report metrics that match the intended scientific use. One step MSE is informative about interpolation quality, but it is insufficient for long horizon plasma rollout because it can reward models that violate charge accounting. Rollout MSE is more useful, yet still incomplete when two models occupy different conservation regimes. Charge budget error, minimum density diagnostics, and raw negativity measurements expose failure modes that are hidden by aggregate state error. For structure preserving neural PDE methods, these quantities should be treated as first class evaluation targets rather than auxiliary sanity checks. The present benchmark is designed around this principle: the evaluation protocol is the core scientific contribution alongside the method, and the baseline without learned correction is a necessary reference for separating structural effects from learned ones.

Finally, the present boundary simplification matters. The experiments isolate conservation, positivity, and Poisson compatibility under zero species wall fluxes and prescribed electrostatic potential boundaries. Real plasma wall interaction introduces collection, emission, secondary electron effects, sheath edge constraints, and kinetic corrections that are not modeled here. The significance of the current result is therefore architectural: it establishes that a conservative learned update can protect the long horizon bookkeeping of a Poisson coupled plasma transport system. A full plasma wall surrogate will need to attach richer, physically justified wall closures to the same conservative backbone rather than replace the backbone with unconstrained state regression.

The comparison should also be read as a first controlled benchmark, not as a final baseline suite. The Poisson recomputed, charge projected, and rollout trained direct baselines show that neither electrostatic compatibility, global charge correction, nor short autoregressive training alone recovers the conservative solver's rollout behavior. However, these are still intermediate checks. Scheduled sampling, stronger projection methods, neural finite volume variants, and operator models with explicit projection could reduce the gap. The current contribution is therefore narrower than a general claim that \method\ dominates learned plasma simulation: it demonstrates that the conservative finite volume inductive bias prevents failure modes that direct state regressors exhibit in the controlled benchmark suite. A stronger paper would need at least one setting where the learned correction clearly improves over the classical core, establishing that the neural component solves a genuine closure problem rather than merely riding on the structural backbone.

\section{Limitations}
This work has several limitations that define its intended scope. The benchmark uses zero species wall flux boundaries with Dirichlet electrostatic potential boundaries, so it is not a full physical sheath wall model with collection, emission, or Bohm like outflow physics. The study is also one dimensional and uses a drift diffusion Poisson fluid description rather than kinetic plasma dynamics; consequently, it does not address multidimensional mesh topology, complex geometry, magnetic effects, or kinetic wall interaction. These restrictions are deliberate for the present controlled benchmark, but they limit the physical claims that can be made from the results.

The evidence also shows that the learned correction is not yet essential. In the headline benchmark, the variant without learned correction outperforms the learned correction on rollout MSE, and the teacher mismatch experiment does not currently show a clear advantage for learning over the conservative core alone. The main learned baseline is intentionally simple, while the added projection and short rollout variants are diagnostic rather than exhaustive. Stronger projection methods, scheduled sampling, neural operator models, and neural finite volume baselines remain necessary for a broader comparative claim. In addition, the data generator is close to the proposed conservative backbone except in synthetic teacher mismatch cases, so similarity between the target solver and the model structure may contribute to the observed advantage.

Positivity and scaling behavior remain important technical boundaries. Positivity handling is imperfect in edge cases: at the highest tested mesh resolution, the raw centered explicit proposal becomes strongly nonadmissible before the limiter and safeguard are applied. The current results also suggest that positivity weight sensitivity is mainly a baseline regularization phenomenon rather than a strong driver of the conservative solver. More broadly, the centered explicit transport scheme becomes genuinely unstable at higher resolution, with raw minimum density $-19.7$ at grid $N_x=64$; Scharfetter Gummel or upwind discretizations are therefore likely necessary for resolution robustness and are not merely optional extensions. The present experiments do not establish resolution transfer without retraining, robustness to noisy data, nonzero wall flux behavior, or computational advantage over the classical solver. Sensitivity to the manually chosen correction scale ($0.05$), the limiter safeguard floor ($\varepsilon=10^{-8}$), CFL type stability limits, and drift dominated centered flux regimes has not been fully mapped.

\section{Conclusion}
We presented a structure preserving rollout study for a controlled one dimensional drift diffusion Poisson benchmark, a PDE system with a long history in plasma transport modeling. The central finding is that conservative discrete structure, not the learned correction, is the dominant stabilizing factor: a classical finite volume core without learned correction already achieves near roundoff rollout error, while the learned flux correction provides an extensible mechanism for future closure settings but does not yet demonstrate a clear standalone advantage. By combining a compatible variable layout, conservative finite volume update, and positivity aware flux handling, the conservative model achieves near roundoff charge accounting by algebraic construction and substantially more stable long horizon rollout than direct next state regression baselines across the benchmark suite. The intermediate baseline comparisons show that neither exact Poisson recomputation, global charge projection, nor short rollout training alone reproduces the local finite volume structure advantage. Across $64$ prespecified benchmark configurations, the conservative structure achieves near roundoff charge error by construction and wins rollout MSE in $60/64$ cases, even though it does not consistently win one step MSE. The main lesson is therefore architectural: for learned plasma transport surrogates, conservative discrete structure is the foundation on which useful learned corrections should be built, and this structural contribution currently dominates the learned one.

Future work should retain this structure first design while adding settings where learning is essential: missing nonlinear closures, nonzero wall flux boundary physics, cross solver generalization, and resolution transfer. The natural next steps are richer wall flux closures attached to the same conservative backbone, Scharfetter Gummel or upwind backbone discretizations for high resolution robustness, stronger learned transport models, and direct comparisons against neural finite volume and operator baselines.

\section*{Data and Code Availability}
The source code, experiment configurations, supporting this study are publicly available at \url{https://github.com/Airscker/DDP-benchmark}. The repository is intended to provide the training and evaluation workflow, manuscript figure generation inputs, and tabulated benchmark outputs needed to reproduce the reported comparisons.

% \section*{Acknowledgments}
% Acknowledgments omitted in this draft version and should be added in the submission ready manuscript.

\bibliographystyle{unsrt}
\bibliography{references}

\clearpage
\appendix
\renewcommand{\thetable}{A\arabic{table}}
\setcounter{table}{0}
\section{Appendix Tables}

\begin{table}[!h]
  \centering
  \caption{Multiseed aggregate (mean $\pm$ standard deviation).}
  \label{tab:multiseed}
  \resizebox{\linewidth}{!}{
  \begin{tabular}{lcccc}
    \toprule
    Model & One step MSE & Rollout MSE & Rollout charge error & Rollout min density \\
    \midrule
    \method & $(1.90 \pm 0.43)\times 10^{-12}$ & $(7.76 \pm 2.10)\times 10^{-9}$ & $(6.76 \pm 1.79)\times 10^{-15}$ & $0.3701 \pm 0.0001$ \\
    \baseline & $(2.80 \pm 0.34)\times 10^{-7}$ & $44.39 \pm 29.50$ & $1.36 \pm 0.98$ & $1.00\times 10^{-14}$ \\
    \bottomrule
  \end{tabular}
  }
\end{table}

\begin{table}[!h]
  \centering
  \caption{Representative harder regime stress tests.}
  \label{tab:stress}
  \resizebox{\linewidth}{!}{
  \begin{tabular}{lcccc}
    \toprule
    Case / Model & One step MSE & Rollout MSE & Rollout charge error & Rollout min density \\
    \midrule
    Bias stronger / \method & $2.03\times 10^{-12}$ & $7.82\times 10^{-9}$ & $4.20\times 10^{-15}$ & $1.11\times 10^{-1}$ \\
    Bias stronger / \baseline & $3.13\times 10^{-7}$ & $1.25\times 10^{1}$ & $1.45\times 10^{0}$ & $1.00\times 10^{-14}$ \\
    Data sparse / \method & $2.45\times 10^{-12}$ & $1.32\times 10^{-8}$ & $4.48\times 10^{-15}$ & $3.70\times 10^{-1}$ \\
    Data sparse / \baseline & $6.81\times 10^{-7}$ & $4.96\times 10^{1}$ & $8.84\times 10^{0}$ & $1.00\times 10^{-14}$ \\
    Stiff transport / \method & $9.52\times 10^{-13}$ & $7.20\times 10^{-9}$ & $6.50\times 10^{-15}$ & $1.60\times 10^{-1}$ \\
    Stiff transport / \baseline & $2.89\times 10^{-7}$ & $5.47\times 10^{1}$ & $9.54\times 10^{-1}$ & $1.00\times 10^{-14}$ \\
    Long horizon sparse / \method & $2.85\times 10^{-12}$ & $1.19\times 10^{-7}$ & $1.60\times 10^{-14}$ & $9.16\times 10^{-2}$ \\
    Long horizon sparse / \baseline & $7.85\times 10^{-7}$ & $4.85\times 10^{2}$ & $1.13\times 10^{2}$ & $1.00\times 10^{-14}$ \\
    \bottomrule
  \end{tabular}
  }
\end{table}

\begin{table}[!h]
  \centering
  \caption{Tradeoff cases where one step or rollout MSE alone would be misleading.}
  \label{tab:tradeoffs}
  \resizebox{\linewidth}{!}{
  \begin{tabular}{lcccc}
    \toprule
    Case / Model & One step MSE & Rollout MSE & Rollout charge error & Rollout min density \\
    \midrule
    Layers 4 / \method & $1.74\times 10^{-2}$ & $4.03\times 10^{-2}$ & $1.08\times 10^{-14}$ & $1.52\times 10^{-1}$ \\
    Layers 4 / \baseline & $5.43\times 10^{-3}$ & $1.36\times 10^{-2}$ & $5.25\times 10^{-3}$ & $3.81\times 10^{-1}$ \\
    Positivity 1.0 / \method & $1.63\times 10^{-2}$ & $2.52\times 10^{0}$ & $4.26\times 10^{-14}$ & $1.00\times 10^{-6}$ \\
    Positivity 1.0 / \baseline & $2.45\times 10^{-3}$ & $4.50\times 10^{-1}$ & $2.61\times 10^{-1}$ & $1.00\times 10^{-14}$ \\
    \bottomrule
  \end{tabular}
  }
\end{table}

\begin{table}[!h]
  \centering
  \caption{Selected generalization and scaling cases.}
  \label{tab:generalization}
  \resizebox{\linewidth}{!}{
  \begin{tabular}{lcccc}
    \toprule
    Case / Model & One step MSE & Rollout MSE & Rollout charge error & Rollout min density \\
    \midrule
    Bias strong / \method & $1.70\times 10^{-2}$ & $7.13\times 10^{-2}$ & $1.44\times 10^{-14}$ & $6.57\times 10^{-2}$ \\
    Bias strong / \baseline & $2.53\times 10^{-2}$ & $3.72\times 10^{0}$ & $2.39\times 10^{0}$ & $1.00\times 10^{-14}$ \\
    4 trajectories / \method & $1.79\times 10^{-2}$ & $4.04\times 10^{-2}$ & $2.09\times 10^{-14}$ & $1.53\times 10^{-1}$ \\
    4 trajectories / \baseline & $1.10\times 10^{-2}$ & $8.76\times 10^{0}$ & $5.61\times 10^{-1}$ & $1.00\times 10^{-14}$ \\
    2 epochs / \method & $1.74\times 10^{-2}$ & $4.00\times 10^{-2}$ & $1.08\times 10^{-14}$ & $1.51\times 10^{-1}$ \\
    2 epochs / \baseline & $1.85\times 10^{-2}$ & $9.91\times 10^{0}$ & $1.25\times 10^{0}$ & $1.00\times 10^{-14}$ \\
    Grid 64 / \method & $1.35\times 10^{-1}$ & $7.04\times 10^{-1}$ & $9.04\times 10^{-15}$ & $1.00\times 10^{-14}$ \\
    Grid 64 / \baseline & $8.44\times 10^{-3}$ & $1.03\times 10^{1}$ & $9.48\times 10^{-2}$ & $1.00\times 10^{-14}$ \\
    \bottomrule
  \end{tabular}
  }
\end{table}

\clearpage
% \color{red}
\scriptsize
\begin{longtable}{@{}p{0.32\linewidth}rrrrrrrr@{}}
\caption{Complete active comparative evaluation registry. Abbreviations: Cons., conservative model; Base., direct baseline; CE, rollout charge error.} \label{tab:all_active_evaluations}\\
\toprule
Case & $N_x$ & Steps & Traj. & Eval & Cons. MSE & Base MSE & Cons. CE & Base CE \\
\midrule
\endfirsthead
\toprule
Case & $N_x$ & Steps & Traj. & Eval & Cons. MSE & Base MSE & Cons. CE & Base CE \\
\midrule
\endhead
1. 10 headline & 16 & 64 & 64 & 128 & 7.35e-09 & 42.3 & 5.93e-15 & 4.48 \\
2. 15 multi seed / seed 001 & 16 & 64 & 64 & 128 & 6.56e-09 & 18.9 & 6.49e-15 & 0.358 \\
3. 15 multi seed / seed 002 & 16 & 64 & 64 & 128 & 6.14e-09 & 58.4 & 3.57e-15 & 2.22 \\
4. 15 multi seed / seed 003 & 16 & 64 & 64 & 128 & 7.19e-09 & 29.1 & 8.75e-15 & 2.72 \\
5. 15 multi seed / seed 004 & 16 & 64 & 64 & 128 & 1.19e-08 & 19.6 & 6.86e-15 & 0.257 \\
6. 15 multi seed / seed 005 & 16 & 64 & 64 & 128 & 7.05e-09 & 95.9 & 8.11e-15 & 1.27 \\
7. 18 harder regimes / bias stronger & 16 & 64 & 64 & 128 & 7.82e-09 & 12.5 & 4.20e-15 & 1.45 \\
8. 18 harder regimes / data sparse & 16 & 64 & 8 & 128 & 1.32e-08 & 49.6 & 4.48e-15 & 8.84 \\
9. 18 harder regimes / long horizon sparse & 16 & 64 & 8 & 256 & 1.19e-07 & 485 & 1.60e-14 & 113 \\
10. 18 harder regimes / stiff transport & 16 & 96 & 32 & 160 & 7.20e-09 & 54.7 & 6.50e-15 & 0.954 \\
11. 19 core ablation / classical core only & 16 & 64 & 64 & 128 & 1.15e-14 & 42.3 & 6.88e-15 & 4.48 \\
12. 19 core ablation / correction scale small & 16 & 64 & 64 & 128 & 2.50e-10 & 42.3 & 1.11e-14 & 4.48 \\
13. 20 stability / rollout 016 & 16 & 64 & 64 & 16 & 2.34e-11 & 0.0122 & 9.92e-16 & 4.86e-04 \\
14. 20 stability / rollout 032 & 16 & 64 & 64 & 32 & 1.36e-10 & 0.276 & 1.89e-15 & 0.0203 \\
15. 20 stability / rollout 064 & 16 & 64 & 64 & 64 & 9.62e-10 & 4.27 & 2.95e-15 & 0.522 \\
16. 20 stability / rollout 128 & 16 & 64 & 64 & 128 & 7.35e-09 & 42.3 & 5.93e-15 & 4.48 \\
17. 23 learned value v2 / teacher flux classical core only v2 & 8 & 64 & 64 & 128 & 0.0406 & 1.56 & 1.09e-14 & 0.77 \\
18. 23 learned value v2 / teacher flux formal v2 & 8 & 64 & 64 & 128 & 0.0404 & 1.56 & 1.63e-14 & 0.77 \\
19. 23 learned value v2 / teacher flux positivity stress classical core only v2 & 8 & 64 & 64 & 128 & 0.212 & 0.96 & 2.23e-14 & 0.436 \\
20. 23 learned value v2 / teacher flux positivity stress formal v2 & 8 & 64 & 64 & 128 & 0.207 & 0.96 & 2.10e-14 & 0.436 \\
21. 31 generalization v2 / bias mild & 8 & 64 & 64 & 128 & 0.0216 & 0.45 & 1.47e-14 & 0.0955 \\
22. 31 generalization v2 / bias strong & 8 & 64 & 64 & 128 & 0.0713 & 3.72 & 1.44e-14 & 2.39 \\
23. 31 generalization v2 / dt large & 8 & 64 & 64 & 128 & 0.0796 & 1.56 & 1.29e-14 & 0.77 \\
24. 31 generalization v2 / dt small & 8 & 64 & 64 & 128 & 0.0146 & 1.58 & 1.70e-14 & 0.77 \\
25. 31 generalization v2 / electron diffusivity high & 8 & 64 & 64 & 128 & 0.0517 & 1.67 & 7.93e-15 & 0.77 \\
26. 31 generalization v2 / electron diffusivity low & 8 & 64 & 64 & 128 & 0.0212 & 1.41 & 1.38e-14 & 0.77 \\
27. 31 generalization v2 / electron mobility high & 8 & 64 & 64 & 128 & 0.0735 & 1.5 & 1.36e-14 & 0.77 \\
28. 31 generalization v2 / electron mobility low & 8 & 64 & 64 & 128 & 0.023 & 1.67 & 8.11e-15 & 0.77 \\
29. 31 generalization v2 / ion diffusivity high & 8 & 64 & 64 & 128 & 0.0423 & 1.57 & 1.03e-14 & 0.77 \\
30. 31 generalization v2 / ion diffusivity low & 8 & 64 & 64 & 128 & 0.0407 & 1.56 & 1.41e-14 & 0.77 \\
31. 31 generalization v2 / ion mobility high & 8 & 64 & 64 & 128 & 0.0923 & 1.55 & 8.35e-15 & 0.77 \\
32. 31 generalization v2 / ion mobility low & 8 & 64 & 64 & 128 & 0.0327 & 1.57 & 2.24e-14 & 0.77 \\
33. 41 mesh resolution v2 / grid 008 & 8 & 64 & 64 & 128 & 0.0404 & 1.56 & 1.63e-14 & 0.77 \\
34. 41 mesh resolution v2 / grid 016 & 16 & 64 & 64 & 128 & 0.119 & 14.2 & 4.04e-15 & 1.48 \\
35. 41 mesh resolution v2 / grid 032 & 32 & 64 & 64 & 128 & 0.459 & 13.4 & 1.32e-14 & 0.245 \\
36. 41 mesh resolution v2 / grid 064 & 64 & 64 & 64 & 128 & 0.704 & 10.3 & 9.04e-15 & 0.0948 \\
37. 51 data regime v2 / trajectories 004 & 8 & 64 & 4 & 128 & 0.0404 & 8.76 & 2.09e-14 & 0.561 \\
38. 51 data regime v2 / trajectories 008 & 8 & 64 & 8 & 128 & 0.0401 & 12.6 & 9.20e-15 & 8.32 \\
39. 51 data regime v2 / trajectories 016 & 8 & 64 & 16 & 128 & 0.0401 & 28.8 & 1.20e-14 & 0.375 \\
40. 51 data regime v2 / trajectories 032 & 8 & 64 & 32 & 128 & 0.0402 & 2.03 & 9.36e-15 & 0.739 \\
41. 51 data regime v2 / trajectories 064 & 8 & 64 & 64 & 128 & 0.0404 & 1.56 & 1.63e-14 & 0.77 \\
42. 61 training duration v2 / epochs 002 & 8 & 64 & 64 & 128 & 0.04 & 9.91 & 1.08e-14 & 1.25 \\
43. 61 training duration v2 / epochs 010 & 8 & 64 & 64 & 128 & 0.0392 & 4.3 & 1.34e-14 & 0.803 \\
44. 61 training duration v2 / epochs 025 & 8 & 64 & 64 & 128 & 0.0405 & 4.04 & 1.42e-14 & 1.67 \\
45. 61 training duration v2 / epochs 050 & 8 & 64 & 64 & 128 & 0.0405 & 2.86 & 1.50e-14 & 1.12 \\
46. 61 training duration v2 / epochs 100 & 8 & 64 & 64 & 128 & 0.0404 & 1.56 & 1.63e-14 & 0.77 \\
47. 71 model capacity v2 / hidden 016 & 8 & 64 & 64 & 128 & 0.039 & 0.0548 & 1.52e-14 & 0.0633 \\
48. 71 model capacity v2 / hidden 032 & 8 & 64 & 64 & 128 & 0.0401 & 0.206 & 1.18e-14 & 0.0467 \\
49. 71 model capacity v2 / hidden 064 & 8 & 64 & 64 & 128 & 0.0404 & 1.56 & 1.63e-14 & 0.77 \\
50. 71 model capacity v2 / layers 2 & 8 & 64 & 64 & 128 & 0.0404 & 0.363 & 1.57e-14 & 0.265 \\
51. 71 model capacity v2 / layers 3 & 8 & 64 & 64 & 128 & 0.0404 & 1.56 & 1.63e-14 & 0.77 \\
52. 71 model capacity v2 / layers 4 & 8 & 64 & 64 & 128 & 0.0403 & 0.0136 & 1.08e-14 & 5.25e-03 \\
53. 81 loss weights v2 / conservation 0p0 & 8 & 64 & 64 & 128 & 0.0404 & 2.43 & 1.01e-14 & 0.582 \\
54. 81 loss weights v2 / conservation 0p1 & 8 & 64 & 64 & 128 & 0.0404 & 1.56 & 1.63e-14 & 0.77 \\
55. 81 loss weights v2 / conservation 10p0 & 8 & 64 & 64 & 128 & 0.0405 & 0.28 & 2.41e-14 & 0.0393 \\
56. 81 loss weights v2 / conservation 1p0 & 8 & 64 & 64 & 128 & 0.0403 & 0.298 & 1.68e-14 & 0.0153 \\
57. 81 loss weights v2 / positivity 0p0 & 8 & 64 & 64 & 128 & 0.214 & 3.91 & 2.28e-14 & 2.95 \\
58. 81 loss weights v2 / positivity 0p1 & 8 & 64 & 64 & 128 & 0.214 & 3.91 & 2.28e-14 & 2.95 \\
59. 81 loss weights v2 / positivity 10p0 & 8 & 64 & 64 & 128 & 0.214 & 3.91 & 2.28e-14 & 2.95 \\
60. 81 loss weights v2 / positivity 1p0 & 8 & 64 & 64 & 128 & 0.214 & 3.91 & 2.28e-14 & 2.95 \\
61. 82 positivity activation v3 / positivity 0p0 & 8 & 64 & 64 & 128 & 2.52 & 0.446 & 4.26e-14 & 0.0206 \\
62. 82 positivity activation v3 / positivity 0p1 & 8 & 64 & 64 & 128 & 2.52 & 0.278 & 4.26e-14 & 0.0124 \\
63. 82 positivity activation v3 / positivity 10p0 & 8 & 64 & 64 & 128 & 2.52 & 5.69 & 4.26e-14 & 4.7 \\
64. 82 positivity activation v3 / positivity 1p0 & 8 & 64 & 64 & 128 & 2.52 & 0.45 & 4.26e-14 & 0.261 \\
\bottomrule
\end{longtable}
\normalsize

\end{document}